\documentclass[aps,prl,twocolumn,superscriptaddress,groupedaddress,amsmath,amssymb,nofootinbib,floatfix]{revtex4}  
\usepackage{graphicx}  
\usepackage{dcolumn}   
\usepackage{bm,color}        
\usepackage{amssymb}   

\newcommand{\be}{\begin{equation}}
\newcommand{\ee}{\end{equation}}
\def\CL{CosmoLib$2^{\rm nd}$} 
\def\rec{*}
\def\D{D_\rec}

\begin{document}

\title{The full CMB temperature bispectrum from single-field inflation}
\author{Zhiqi Huang}
\address{CITA, University of Toronto, 60 St. George Street, Toronto, Ontario M5S 3H8, Canada}
         \author{Filippo Vernizzi}
\address{CEA, Institut de Physique Th{\'e}orique,
         F-91191 Gif-sur-Yvette c{\'e}dex,  
         CNRS, Unit{\'e} de recherche associ{\'e}e-2306, F-91191 Gif-sur-Yvette c{\'e}dex}
\date{\today}


\begin{abstract}
We compute the full cosmic microwave background temperature bispectrum generated by nonlinearities after single-field inflation. 
By integrating the photon temperature at second order along a perturbed geodesic in Newtonian gauge, we derive an expression for the observed temperature fluctuations that, for the first time, clarifies  the separation of the gravitational lensing and time-delay effects from the purely second-order contributions. 
We then use the second-order  Boltzmann code  CosmoLib$2^{\rm nd}$ to calculate these contributions and their  bispectrum. Including the perturbations in the photon path, the numerically computed bispectrum  exactly matches the expected squeezed limit. 
Moreover, the analytic squeezed-limit formula  reproduces well the signal-to-noise and shape of the full bispectrum,
potentially facilitating the subtraction of the bias induced by second-order effects. For a cosmic-variance limited experiment with $l_{\rm max} = 2000$, the bias on a local signal is $f_{\rm NL}^{\rm loc} =0.73$ negligible for  equilateral and orthogonal  signals.
The signal-to-noise ratio is unity at $l_{\rm max} \sim 3000$, suggesting that second-order effects may  hopefully be measured in the future.
\end{abstract}

\maketitle

One of the main results of the Planck satellite nominal mission is that primordial non-Gaussianity is small \cite{Ade:2013ydc}. It is tempting to conclude that single-field models of inflation are confirmed (although multi-field inflation is still compatible with small non-Gaussianity \cite{Vernizzi:2006ve}); however, the current constraints still allow for many models predicting $f_{\rm NL}^{\rm loc} \sim $ few, which may be detected once that all the 2.5 years temperature and polarization data will be considered.
Since so much is at stake, any little improvement on these constraints may very important.

To clean the data from any contamination of the primordial signal we must consistently account for the nonlinear relation between the initial conditions and the CMB anisotropies. 
An important nonlinear effect that has been clearly detected \cite{Ade:2013ydc} comes from the integrated Sachs-Wolfe (ISW)-lensing correlation  (see e.g.~\cite{Lewis:2012tc}). 

Other late-time nonlinearities are expected to contribute,  in a minor part, to the bispectrum. Although small, their subtraction should be taken into account.  
Due to the complexity of their calculation, these effects require dedicated numerical studies. During the last few years there has been an intense effort to derive the complete second-order equations \cite{Bartolo:2004ty,Pitrou:2008hy,Khatri:2008kb,Senatore:2008vi,Nitta:2009jp,Boubekeur:2009uk,Pitrou:2010sn,Beneke:2010eg}, which recently led to the development of \CL~\cite{Huang:2012ub},  a numerical  Boltzmann code at second-order  to compute the CMB bispectrum from nonlinear effects. This code predicted that the bias on local non-Gaussianity is small but non-negligible, which has been  qualitatively confirmed by other two independent numerical studies \cite{Su:2012gt,Pettinari:2013he}.

A challenging aspect of second-order codes is the integration of the photon temperature along the line of sight. In particular, the photon Boltzmann equation contains second-order couplings between the gravitational potentials and {\em all} multipoles of the temperature generated by  free-streaming after recombination \cite{Huang:2012ub}. Since the solution fails to converge by naively truncating at finite $l$, incorrect treatment of these terms has led to overestimating their effect on the bispectrum \cite{Pitrou:2010sn}. In  \cite{Huang:2012ub} we have shown that the solution to this problem is to rewrite the infinite sum of multipoles as a boundary term and a term that contributes only before recombination, where  frequent Thomson scatterings suppress high multipoles. 

On the other hand, all previous calculations have been performed  along an {\em unperturbed} geodesic in Newtonian conformal gauge. However, to ensure that the final result is observable and gauge invariant one needs to include deviations from a straight geodesic and understand the separation of  lensing and time delay  from the other second-order effects. 
As pointed out in \cite{Pettinari:2013he,Naruko:2013aaa}, a longstanding obstacle hindering this inclusion is a convergence problem similar to the one discussed above.

In this Letter, we first remove this obstacle and---to our knowledge for the first time---clarify the separation between lensing, time-delay and intrinsic second-order effects that need a dedicated code to be computed.
Second, we  calculate the {\em full observable} CMB temperature bispectrum (in the absence of primordial non-Gaussianity) and show that only once the above separation is clear one can reach an exact agreement with the analytic formula in the squeezed limit \cite{Creminelli:2004pv,Creminelli:2011sq,Bartolo:2011wb,Lewis:2012tc,Pajer:2013ana}. Finally, we  compare the signal-to-noise of the reduced bispectrum computed with the code to the one  from the squeezed-limit approximation and recompute the values of the contamination up to $l_{\rm max} \sim 3000$ with higher accuracy than previous computations.

We have improved  \CL~with respect to the version used in \cite{Huang:2012ub}. In particular, we now compute fluctuations in the free-electron density by consistently perturbing RECFAST \cite{Seager:1999bc} to first order, thus including Helium recombination; we have also increased the accuracy of the 3-d integrator and consistently included vector and tensor (i.e., $m=\pm 1, \pm2$) contributions. None of these improvements sensibly affects the results of  \cite{Huang:2012ub}. Resonable convergence of the line-of-sight integral can be obtained for $l \le3$ in the second-order source but here we safely use $l \le 7$. We employ Planck cosmological parameters \cite{Ade:2013zuv} (without reionization) in all calculations. We will give more details on the present version of the code in a longer paper \cite{ZF}.

\vskip.1cm
{\em  Full bispectrum.} 
The Boltzmann equation for photons can be written in terms of the fractional brightness $\Delta \equiv \delta I/\bar I$, with $I (\eta,\vec x, \hat n) \equiv \int dp \, p^3 f (\eta, \vec x, \vec p) $; $f$ is the photon  distribution and we have rewritten the momentum of photons in the local inertial frame,  $p^i$, as $p^i \equiv p n^i$, $n_i n^i=1$. Using a perturbed metric at second order in the Poisson gauge, 
$ds^2 = a^2(\eta) [-  e^{2 \Phi}  d\eta^2+2  \omega_i d \eta dx^i +(e^{-2 \Psi} \delta_{ij} + \chi_{ij})dx^i dx^j ]$,
with $\omega_{i,i}=0$ and $\chi_{ii}=0=\chi_{ij,j}$, the brightness equation reads
\be
\label{brightness_eq}
\begin{split}
\frac{d}{d \eta} (\Delta + 4 \Phi)  & \equiv \left( \frac{\partial}{\partial \eta} + \frac{d x^i}{d \eta} \partial_i + \frac{d n^i}{d \eta} \partial_{n^i} \right) (\Delta + 4 \Phi)  \\
&=
  4 \Delta (\dot \Psi - \Phi_{,i} n^i) + E 
    -  ( \dot \tau + \delta \dot \tau ) F\;, 
\end{split}
\ee
where a dot denotes the derivative with respect to $\eta$; $\dot \tau \equiv - n_e \sigma_T a$ is the unperturbed differential optical depth and $\delta \dot \tau \equiv \dot \tau (\delta_e + \Phi)$ its perturbation, where $\delta_e \equiv \delta n_e/n_e$ is the free-electron density contrast. We assume there is no reionization; hence, both $\dot \tau$ and $\delta \dot \tau$ vanish today. In the second line, $E \equiv  4 (\dot \Phi + \dot \Psi) - 4 \dot \omega_i n^i - 2 \dot \chi_{ij} n^i n^j $ is the redshift in  photon energy due to integrated effects: the ISW contribution (whose second-order part is the Rees-Sciama (RS) effect), and the  vector and tensor contributions, respectively \cite{Boubekeur:2009uk}. Finally, the collision term of the Boltzmann equation, $F$, can be read off from the RHS~of eq.~(78) of \cite{Senatore:2008vi}, with the notation for $\Phi$ and $\Psi$ interchanged. 

As explained in \cite{Huang:2012ub}, the first term in the second line of eq.~\eqref{brightness_eq} couples the gravitational potentials to   all the multipole moments of $\Delta$ generated by photon free-streaming along the line of sight.  Hence, solving this equation by naively truncating the multipole expansion at finite order leads to a lack of convergence of its solution. As shown in \cite{Huang:2012ub}, this term can be traded by a total derivative and  
terms proportional to $\dot \tau$, which vanish after recombination and whose multipole expansion can be thus safely truncated at low $l$.
Indeed,  by replacing $\Phi_{,i} n^i $ by $ d \Phi/d\eta - \dot \Phi$ and dividing eq.~\eqref{brightness_eq} by $1+\Delta$ we can rewrite it, up to second order, as
\begin{align}
\label{brightness_eq2}
&\frac{d}{d \eta} \left[ \left( \Delta_G + 4 \Phi \right)  e^{ - (\tau + \delta \tau) } \right]  = (E  -   \dot \tau  R) e^{ -(\tau + \delta \tau) }  \;, \\
&\dot \tau R \equiv (\dot \tau + \delta \dot \tau) ( F_G +  \Delta_G + 4 \Phi) \;,
\end{align}
with $\Delta_G \equiv \Delta - \frac12 \Delta^2$ and $F_G \equiv F (1-\Delta)$.

We want to integrate
this equation along the perturbed photon trajectory. 
While at linear order one usually relies on the so called Born approximation and integrate the first-order source along an unperturbed geodesic, here we need to go at one order higher. 
To do that, we define the lensing deviation  $\delta x^i \equiv \vec x (\eta, \hat n) - \vec x_0 (\eta) $, where $\vec x_0(\eta)\equiv \hat n(\eta - \eta_0)$ is the unperturbed   geodesic,  and the deviation angle $\delta n^i$, with $\delta n^i(\eta_0)=0$. The evolution of these quantities along the line of sight can be computed using the photon geodesic equation (see e.g.~eqs.~(63) and (65) of \cite{Senatore:2008vi}), 
\begin{align}
{d \delta x^i}/{d \eta}  &=  \delta n^i +  n^i (\Phi+\Psi)\;,  \label{cv}\\
{d \delta n^i}/{d \eta} & =  - \nabla_{\perp}^i (\Phi +\Psi) \label{ne}  \;,
\end{align}
with $\nabla_{\perp}^i \equiv (\delta^{ij} - n^i n^j) \partial_j$.
Plugging these equations into the definition of  convective derivative in the first line of eq.~\eqref{brightness_eq}, we can formally solve eq.~\eqref{brightness_eq2} as
\be
\label{int_eq_source}
\begin{split}
&  \Delta_{\rm obs}( \hat n) = \frac12 \left[ \Delta_{\rm obs} ( \hat n) \right]^2 + \int_0^{\eta_0} \! \! d \eta e^{-\tau}   \big\{ (1  - \delta \tau)  (E  -   \dot \tau  R) \\
  &-
 \left[ (\Phi + \Psi)n^i  \partial_i   -  \nabla_{\perp}^i (\Phi +\Psi) \partial_{n^i} \right]  (\Delta + 4 \Phi)  \big\},
\end{split}
\ee
where now  all the quantities in the integrand  on the RHS are evaluated 
along the {\em unperturbed  } geodesic.

In Ref.~\cite{Huang:2012ub}, the  last line of this equation was not included in the source for the calculation of the bispectrum.
This line seems to involve the same difficulties encountered in the calculation of the first term in the second line of eq.~\eqref{brightness_eq}, i.e.~couplings between $\Phi$ and $\Psi$ with the full hierarchy of multipoles of  $\Delta$. 
However, also in this case we can rewrite it as a  total derivative---yielding a boundary term once integrated---and terms which can be truncated at finite $l$. Indeed, after an integration by parts $-e^{-\tau} (\Phi + \Psi)n^i  \partial_i (\Delta + 4 \Phi)$ becomes  
\be
\label{time_delay}
 \frac{d}{d \eta} \left[ \phi \D n^i  \partial_i  (\Delta+ 4 \Phi) \right]  - \D  \phi \,  n^i \partial_i \left( E - \dot \tau  F \right) \;,
\ee
where we have defined the gravitational {\em time-delay potential} at time $\eta$   and the angular diameter distance to recombination, respectively as
\begin{align}
\label{td}
\phi (\eta ,\hat n ) &\equiv  - \frac{1}{\D} \int_0^{\eta} d \eta' e^{-\tau} (\Phi+\Psi) \;, \\
\D&\equiv \eta_0 - \eta_\rec \;, \qquad \eta_\rec \equiv    \int_0^{\eta_0} d \eta \dot \tau e^{-\tau} \eta   \;.
\end{align}
It is lengthier but  straightforward to show that $e^{-\tau} \nabla_{\perp}^i (\Phi +\Psi) \partial_{n^i} (\Delta + 4 \Phi)$ can be rewritten as
\be
\label{Delta_lensing}
\begin{split}
&\frac{d}{d \eta}  \left[ \partial_{n^i} \psi   \partial_{n^i} ( \Delta+ 4 \Phi)  \right] -  \partial_{n^i} \psi \frac{d}{d n^i} \left( E - \dot \tau  F \right) \\
&+ \frac{d \partial_{n^i} \psi}{d \eta} (\eta -\eta_0)  e^{-\tau} \dot \tau \partial_{n^i} R
\;,
\end{split}
\ee
where $\frac{d}{d n^i}  \equiv  \partial_{n^i} + (\eta - \eta_0) \partial_i $ and we have defined 
 the gravitational {\em lensing potential} at time $\eta$  as
\begin{align}
\label{alpha}
\psi (\eta,   \hat n) &\equiv \int_0^{\eta_0} d\eta' g \left(\min(\eta,\eta') \right) (\Phi+ \Psi) \;, \\
g (\eta) &\equiv \frac1{\eta-\eta_0} \int_0^{\eta} d \eta'  \dot \tau e^{-\tau} \frac{ \eta-\eta'}{\eta'-\eta_0} \;,
\end{align}
which shows that the lensing angle $\partial_{n^i} \psi$ describes the difference between the  deviation angle at time $\eta$ and that at the time of last scattering.
The photon brightness appearing in $F$ and $R$ in eqs.~\eqref{time_delay} and \eqref{Delta_lensing} is always proportional to $\dot \tau$; hence, higher-order multipoles are  suppressed and we can safely truncate the multipole expansion of $\Delta$ at finite $l$.
The second line in eq.~\eqref{Delta_lensing} takes into account that sources at the last scattering are generally anisotropic and gravitational lensing changes the angle at which they are viewed. This contribution only affects the very small multipoles  \cite{Hu:2001yq} and we neglect it here.

In conclusion, replacing the last line of eq.~\eqref{int_eq_source} with the expressions \eqref{time_delay} and \eqref{Delta_lensing} and integrating the boundary terms, the {\em observed total} CMB temperature anisotropy at second order is given by the sum of four contributions (which are not separately gauge invariant \cite{Creminelli:2004pv}), 
\be
\label{ani_total}
\begin{split}
&\Delta^{(2)}_{\rm obs}(\hat n) =    \frac12 \left[ \Delta_{\rm obs}(\hat n) \right]^2 + \Delta_S^{(2)} (\eta_0 ,\hat n)  \\
&+  \phi (\eta_0,\hat n) \D n^i  \partial_i  \Delta_{\rm obs}(\hat n) + \partial_{n^i} \psi (\eta_0,\hat n)  \partial_{n^i}  \Delta_{\rm obs}(\hat n) \;,
\end{split}
\ee
where $\Delta_S (\eta_0,\hat n)$ is given, up to second order, by
\be
\label{final_Theta}
\begin{split}
& \Delta_S (\eta_0,\hat n)  =   \int_{0}^{\eta_0} d \eta  \bigg[e^{-\tau} (1- \delta \tau) \left( E - \dot \tau  R \right)    \\  & - \D \phi \,  n^i \partial_i \left( E - \dot \tau  F \right)  - \partial_{n^i} \psi \frac{d }{d n^i}  \left( E - \dot \tau  F \right)  \bigg] \;.
\end{split}
\ee
The first term on the RHS of eq.~\eqref{ani_total} comes from the local relation between $\Delta_G$ and $\Delta$ \cite{Huang:2012ub}. The third term is the standard gravitational time delay and the fourth one is lensing. In particular, the potentials $ \phi (\eta_0 ,\hat n )$ and $\psi (\eta_0,\hat n)$  respectively correspond to $ d(\hat n) $  and $ \phi (\hat n)$,  defined in eqs.~(6), (1) and (2) of Ref.~\cite{Hu:2001yq}.     The  time delay is suppressed by $\eta_*/D_* $ and can be  neglected \cite{Hu:2001yq}. Hence the total bispectrum reads 
\be
\label{eq_lensing_rs}
b_{l_1 l_2 l_3} = \big[ ( C_{l_1}  + L_{l_1 l_2 l_3} C_{l_1}^{T \psi} )C_{l_2} + \text{perms} \big]+b_{S,l_1 l_2 l_3}    \;,
\ee
where $L_{l_1 l_2 l_3} \equiv [ l_1(l_1+1) + l_2(l_2+1) - l_3 (l_3+1) ]/2$, $C_l^{T\psi}$ is the cross-correlation spectrum between the temperature and the lensing potential (dominated by the ISW-lensing correlation) and $b_{S,l_1 l_2 l_3}$ is the bispectrum computed from eq.~\eqref{final_Theta}. Since the terms in bracket on the RHS are due to the modulation of the short-scale power spectrum by the long modes, this formula is nonperturbative in the short modes and involve the {\em lensed} small-scale power spectrum rather than the unlensed one \cite{Lewis:2012tc}.
The last term, $b_{S,l_1 l_2 l_3}$, is the only one that requires a second-order Boltzmann calculation. In the rest of this Letter we concentrate our study on this contribution.

\vskip.1cm
{\em  Squeezed limit.} 
As argued in \cite{Creminelli:2011sq,Huang:2012ub}, an important check of any Boltzmann code is to reproduce the bispectrum 
in the squeezed limit, which can be computed analytically because 
it is dominated by the angular modulation of the small-scale power spectrum by super-horizon modes at recombination   \cite{Creminelli:2004pv,Creminelli:2011sq,Bartolo:2011wb,Lewis:2012tc,Pajer:2013ana}. Here we show that the last line of eq.~\eqref{final_Theta} is crucial to correctly reproduce the squeezed-limit formula.

The contribution to the bispectrum from $\Delta_{S}$ in eq.~\eqref{final_Theta} can be computed analytically in the squeezed limit,  by considering the effect of spatial coordinates redefinition at recombination by a long wavelength of the primordial curvature perturbation  $\zeta$, $\vec x \to \vec x (1+\zeta)$. This transformation (which leaves unaffected the lensing angle $\partial_{n^i} \psi$) generates the following contributions to the integrand of eq.~\eqref{final_Theta},
\be
\label{source_trans}
 \zeta (\eta - \eta_0) n^i \partial_i  (\tilde E -\dot \tau R)- \left( \int_0^{\eta} d \eta' e^{-\tau} \zeta \right)  n^i \partial_i \big( \tilde  E - \dot \tau  F \big) \;,
\ee
where we have used $\Phi+\Psi = - \zeta$ for the long wavelength time-delay potential. Since the long mode only modulates quantities at recombination, we use a tilde to denote that we have removed the late-ISW contribution. 
Integrating both terms in eq.~\eqref{source_trans} by parts along the line of sight gives a spatial redefinition of the temperature fluctuation, $\Delta_S^{(2)} (\eta_0 ,\hat n) \approx \zeta \D \, n^i \partial_i  \tilde \Delta (\eta_0 ,\hat n)$. 
In the squeezed limit $l_1 \ll l_2,l_3$ and for $l_1\ll l_H$, where $l_H \simeq (Ha/c_s)_\rec \D   \simeq 110$, this corresponds to the reduced bispectrum
\be
\label{app}
b_{S,l_1 l_2 l_3} \approx    C^{T \zeta }_{l_1}   \frac{1}{l^2 } \frac{ d (l^2  \tilde C_{l})}{d \ln l}  \;,
\ee
where $ \vec l \equiv (\vec l_2 - \vec l_3)/2$ and $C_{l_1}^{T\zeta}$ is the cross-correlation spectrum between the temperature and $\zeta$.
One can verify that a long wavelength time-coordinate transformation $\eta \to \eta + \epsilon(\eta)$ induced by the long mode leaves unchanged eq.~\eqref{final_Theta}  up to corrections of order $\eta_*/D_*$.

\begin{figure}[h]
\begin{center}
{\includegraphics[scale=0.46]{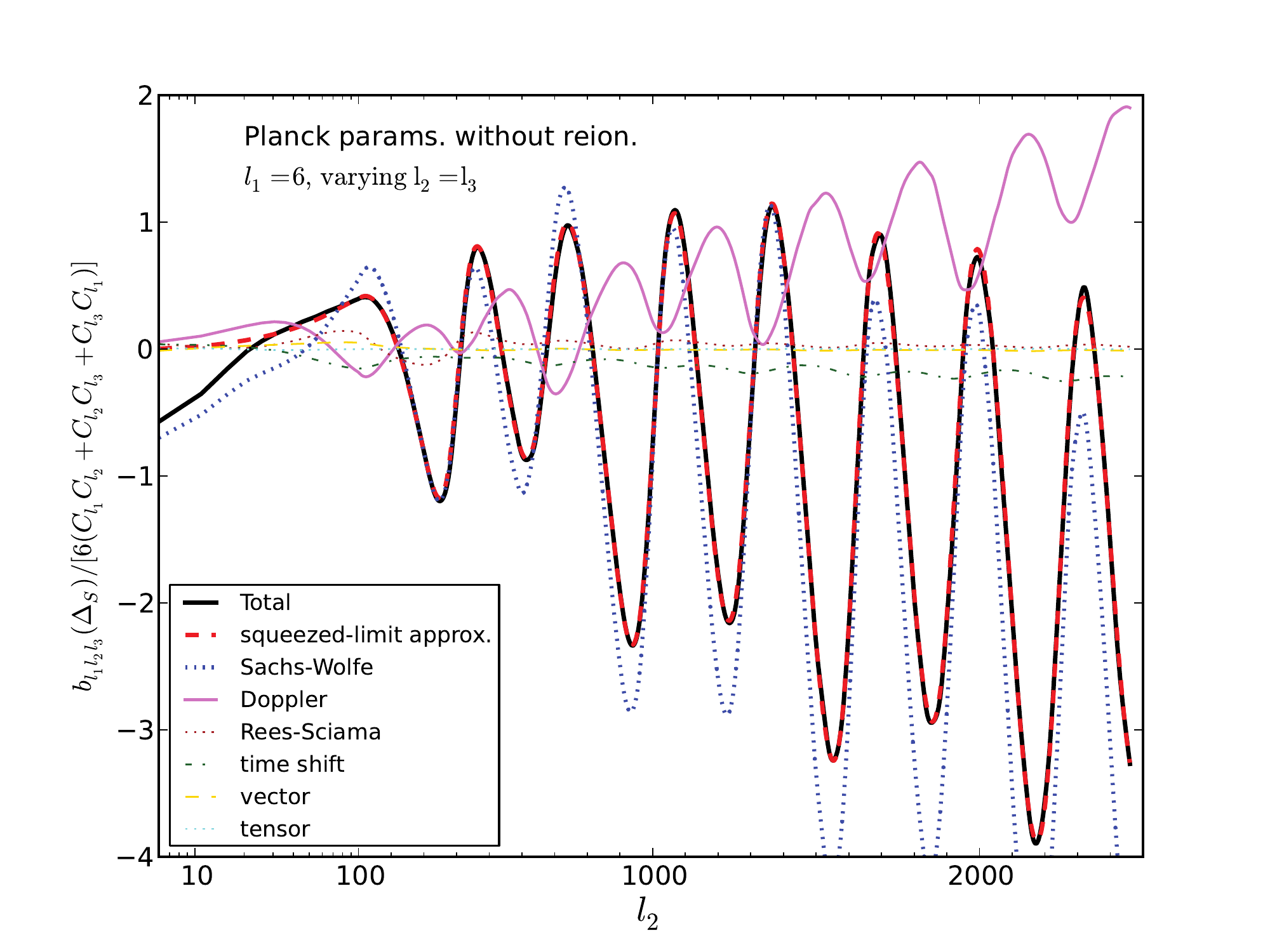}}
\caption{Reduced bispectrum from eq.~\eqref{final_Theta}, $b_{S,l_1 l_2 l_3}$, for $l_1=6$ as a function of $l_2=l_3$, decomposed in its different contributions and compared to the squeezed-limit formula. The ``time shift'' denotes the contribution from the second term on the RHS of eq.~\eqref{final_Theta} while the one from the last term is not shown here due to its smallness in the squeezed limit. The horizontal axis uses logarithmic (linear) scale for $l\le 100$ ($l > 100$).  }
\label{fig:1_l_fixed}
\end{center}
\end{figure} 
In Fig.~\ref{fig:1_l_fixed} we compare the bispectrum numerically computed using eq.~\eqref{final_Theta} with the  approximate formula eq.~\eqref{app}. As expected, the agreement is of ${\cal O} (l_1/l)^2$. While the last term in bracket in eq.~\eqref{final_Theta} can be neglected in the squeezed limit, the third term, here called ``time shift'', becomes important---and crucial to perfectly match the approximate formula---when the early-ISW effect is large. (When $\Lambda=0$ the early-ISW effect is negligible and agreement can be obtained also in the absence of the last line of eq.~\eqref{final_Theta} \cite{Huang:2012ub}.)

\vskip.1cm
{\em  Bispectrum amplitude and shape.} 
We define the Fisher matrix between two bispectra $X$ and $Y$ as \cite{Komatsu:2001rj}
\be
F_{X,Y}  \equiv \sum_{2 \le l_1 \le l_2 \le l_3 \le l_{\rm max}} \frac{B^{(X)}_{l_1l_2l_3} B^{(Y)}_{l_1l_2l_3}}{C_{l_1} C_{l_2} C_{l_3} \Delta_{l_1 l_2 l_3}}\;,
\ee
with
$\Delta_{l_1 l_2 l_3}=1,2,6$ for triangles with no, two or three equal sides. The indices $X$ and $Y$ run from ``sec,  app, loc, eq, ort'', respectively denoting the contributions from second-order effects, i.e.~$C_{l_1} C_{l_2} + C_{l_1} C_{l_3} + C_{l_2} C_{l_3} + b_{S, l_1 l_2 l_3}$, their analytic approximation in the squeezed limit, i.e.~$2 C_{l_1}  C_{l}$ plus eq.~\eqref{app},  the local, equilateral and orthogonal shapes. A complete list of the elements of $F_{X,Y}$ (including the separation with late-time and $m=0,\pm1,\pm2$ contributions) as a function of $l_{\rm max}$ can be found in \cite{webF}. 

\begin{figure}[t]
\begin{center}
{\includegraphics[scale=0.46]{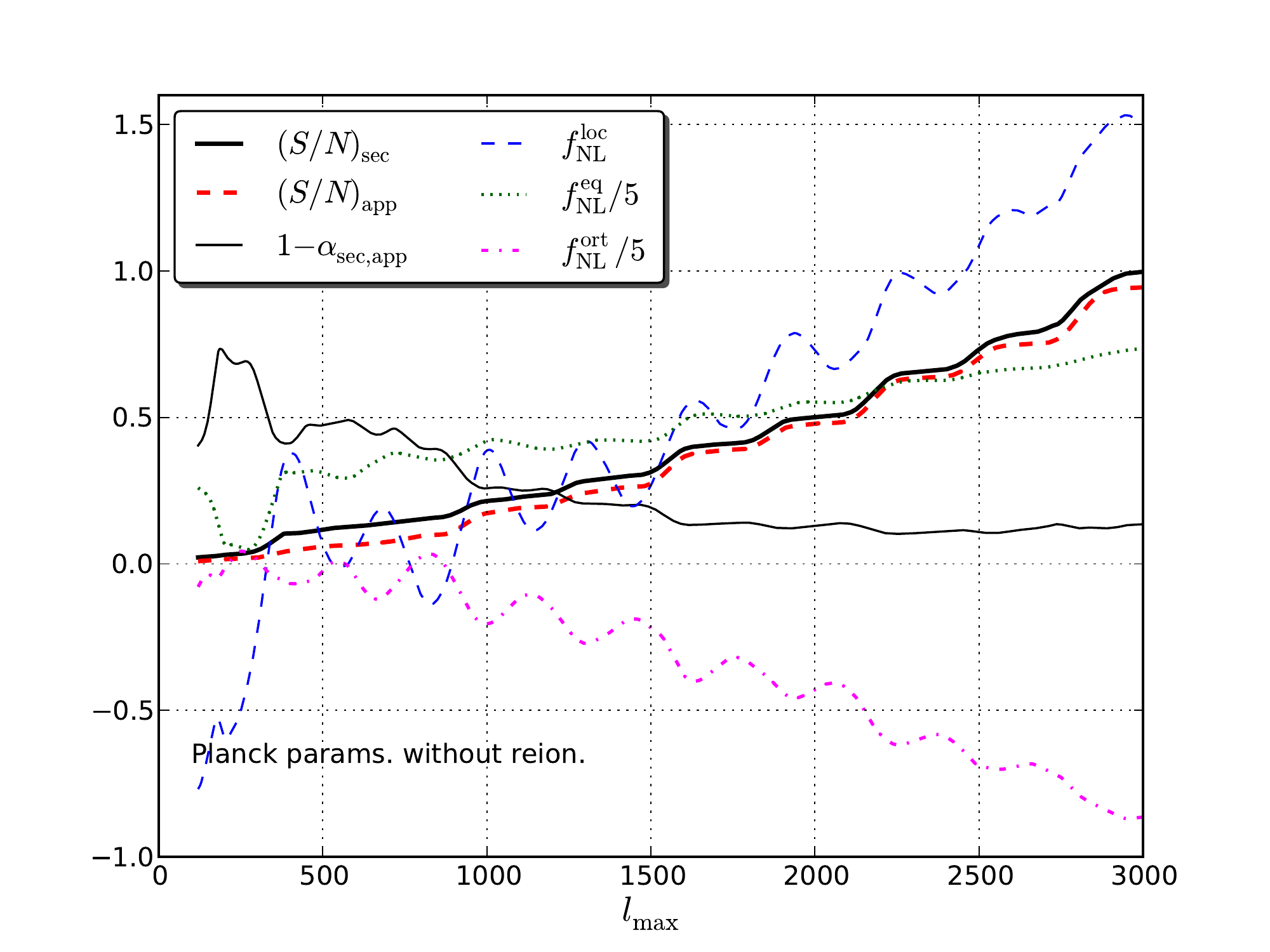}}
\caption{Signal-to-noise ratio of second-order effects,  i.e.~$C_{l_1} C_{l_2} + C_{l_1} C_{l_3} + C_{l_2} C_{l_3} + b_{S, l_1 l_2 l_3}$, compared to the one from the squeezed-limit approximation,  i.e.~$2 C_{l_1}  C_{l}$ plus eq.~\eqref{app}, and bias to local, equilateral and orthogonal signals.}
\label{fig:s2n_all}
\end{center}
\end{figure} 
In Fig.~\ref{fig:s2n_all} we plot the signal-to-noise ratio of second-order effects, $(S/N)_{\rm sec} \equiv F_{\text{sec,sec}}^{1/2}$,  the squeezed-limit approximation, $(S/N)_{\rm app} \equiv F_{\text{app,app}}^{1/2}$, and their overlap defined as $\alpha_{\text{sec,app}} \equiv F_{\rm sec,app}/({F_{\rm sec,sec} F_{\rm app,app}})^{1/2}$. The signal-to-noise agrees  with Ref.~\cite{Pettinari:2013he} (and qualitatively with \cite{Su:2012gt} at $l_{\rm max} = 2000$) and is well approximated, up to $\sim10\%$ differences both in the amplitude and in the shape, by its analytic approximation in the squeezed limit. We have also checked that late-time second-order effects, the most relevant of which are the RS and the late-time part of the time shift, only contribute by $\sim 10\%$ to $(S/N)_{\rm sec}$, roughly independently of $l_{\rm max}$ for  $l_{\rm max} \gtrsim 700$ \cite{webF}. Hence, most of $(S/N)_{\rm sec}$ originates at recombination. Figure~\ref{fig:s2n_all} also shows the bias of second-order sources  ($f^{(X)}_{\rm NL} \equiv F_{X, \rm sec}/F_{X,X}$) on local, equilateral and orthogonal primordial signals. The bias on a local signal is small but non-negligible and should be subtracted from the current constraints. 
Biases on equilateral and orthogonal signals are always smaller than an order of magnitude of their variance and can thus be totally neglected. 
We disagree by $10\%$ with the value of $f_{\rm NL}^{\rm loc}$ reported in \cite{Huang:2012ub}; this is  mainly due to a sub-optimal binning scheme, here corrected \cite{webF}. Comparison with \cite{Su:2012gt,Pettinari:2013he} on the values of the contamination is not straightforward, because these references  integrate different terms along the line of sight.

\vskip.1cm
{\em  Conclusion.} Second-order effects in the CMB temperature are finally completely under control. We have clearly separated  second-order sources at recombination from the better-known ISW-lensing correlation.
Even though the former is  smaller than the latter, we propose that both should be included in future Planck analysis of non-Gaussianity. This task could be  simplified by directly using eq.~\eqref{eq_lensing_rs} with $b_{S,l_1 l_2 l_3}$ given by the analytic formula in the squeezed limit, eq.~\eqref{app} \cite{Creminelli:2011sq,Lewis:2012tc,Pajer:2013ana},
which we  showed to approximate well   the second-order effects. 
 
As the signal-to-noise ratio of these effects becomes  unity at $l_{\rm max} \sim 3000$, they may eventually be measured by combining present and future CMB temperature and polarization  data and become an important cross-check of standard cosmology beyond linear theory.

\vskip.1cm
\emph{N.B.} We thank the authors of \cite{Su:2012gt,Pettinari:2013he}, P.~Creminelli, A.~Lewis,
 C.~Pitrou and B. van Tent
for very stimulating discussions and useful correspondence,  the SNS of Pisa for kind hospitality  and we acknowledge support by the ANR {\it Chaire d'excellence Junior} CMBsecond ANR-09-CEXC-004-01.




\begin{thebibliography}{99}  

\bibitem{Ade:2013ydc} 
  P.~A.~R.~Ade {\it et al.}  [Planck Collaboration],
  arXiv:1303.5084.
  
\bibitem{Vernizzi:2006ve} 
  F.~Vernizzi and D.~Wands,
  JCAP {\bf 0605}, 019 (2006),
  astro-ph/0603799.
    
 \bibitem{Lewis:2012tc} 
  A.~Lewis,
  JCAP {\bf 1206}, 023 (2012),
  arXiv:1204.5018.




\bibitem{Bartolo:2004ty}
  N.~Bartolo, S.~Matarrese and A.~Riotto,
  Phys.\ Rev.\ Lett.\  {\bf 93}, 231301 (2004),
  astro-ph/0407505. {\em Ibid,}
  JCAP {\bf 0606}, 024 (2006),
  astro-ph/0604416. {\em Ibid,}
%
  JCAP {\bf 0701}, 019 (2007),
  astro-ph/0610110. 

    
\bibitem{Pitrou:2008hy}
  C.~Pitrou,
  Class.\ Quant.\ Grav.\  {\bf 26}, 065006 (2009),
  arXiv:0809.3036.


\bibitem{Khatri:2008kb}
  R.~Khatri, B.~D.~Wandelt,
  Phys.\ Rev.\  {\bf D79}, 023501 (2009).,
  arXiv:0810.4370. {\em Ibid,}
  Phys.\ Rev.\  {\bf D81}, 103518 (2010), arXiv:0903.0871.

\bibitem{Senatore:2008vi}
  L.~Senatore, S.~Tassev, M.~Zaldarriaga,
  JCAP {\bf 0908}, 031 (2009),
  arXiv:0812.3652. {\em Ibid,}
  JCAP {\bf 0909}, 038 (2009),
  arXiv:0812.3658.

\bibitem{Nitta:2009jp} 
  D.~Nitta, E.~Komatsu, N.~Bartolo, S.~Matarrese and A.~Riotto,
  JCAP {\bf 0905}, 014 (2009), arXiv:0903.0894.

\bibitem{Boubekeur:2009uk}
  L.~Boubekeur, P.~Creminelli, G.~D'Amico, J.~Norena and F.~Vernizzi,
  JCAP {\bf 0908}, 029 (2009),
  arXiv:0906.0980.

\bibitem{Pitrou:2010sn}
  C.~Pitrou, J.~-P.~Uzan, F.~Bernardeau,
  JCAP {\bf 1007}, 003 (2010),
  arXiv:1003.0481.
  
\bibitem{Beneke:2010eg} 
  M.~Beneke and C.~Fidler,
  Phys.\ Rev.\ D {\bf 82}, 063509 (2010),
  arXiv:1003.1834.

\bibitem{Huang:2012ub} 
  Z.~Huang and F.~Vernizzi,
  Phys.\ Rev.\ Lett.\  {\bf 110}, 101303 (2013),
  arXiv:1212.3573.

  
    \bibitem{Su:2012gt} 
  S.-C.~Su, E.~A.~Lim and E.~P.~S.~Shellard,
  arXiv:1212.6968.
  
\bibitem{Pettinari:2013he} 
  G.~W.~Pettinari, C.~Fidler, R.~Crittenden, K.~Koyama and D.~Wands,
  JCAP {\bf 1304}, 003 (2013),
  arXiv:1302.0832.
  
\bibitem{Creminelli:2011sq} 
  P.~Creminelli, C.~Pitrou and F.~Vernizzi,
  JCAP {\bf 1111}, 025 (2011),
  arXiv:1109.1822.
  
\bibitem{Creminelli:2004pv}
  P.~Creminelli and M.~Zaldarriaga,
  Phys.\ Rev.\  D {\bf 70}, 083532 (2004),
  arXiv:astro-ph/040542.
  
\bibitem{Bartolo:2011wb} 
  N.~Bartolo, S.~Matarrese and A.~Riotto,
  JCAP {\bf 1202}, 017 (2012),
  arXiv:1109.2043.

\bibitem{Pajer:2013ana} 
  E.~Pajer, F.~Schmidt and M.~Zaldarriaga,
  arXiv:1305.0824 [astro-ph.CO].
  

\bibitem{Naruko:2013aaa} 
  A.~Naruko, C.~Pitrou, K.~Koyama and M.~Sasaki,
  Class.\ Quant.\ Grav.\  {\bf 30}, 165008 (2013),
  arXiv:1304.6929.

\bibitem{Seager:1999bc} 
  S.~Seager, D.~D.~Sasselov and D.~Scott,
  Astrophys.\ J.\  {\bf 523}, L1 (1999),
  astro-ph/9909275.
  
\bibitem{Ade:2013zuv} 
  P.~A.~R.~Ade {\it et al.}  [Planck Collaboration],
  arXiv:1303.5076.
  
  \bibitem{ZF} 
  Z.~Huang, F.~Vernizzi,
in preparation.
  
\bibitem{Hu:2001yq}
  W.~Hu, A.~Cooray,
  Phys.\ Rev.\  {\bf D63}, 023504 (2001),
  astro-ph/0008001.


\bibitem{webF} 
\url{ http://www.cita.utoronto.ca/~zqhuang/CosmoLib/fisher.dat}
 

\bibitem{Komatsu:2001rj} 
  E.~Komatsu and D.~N.~Spergel,
  Phys.\ Rev.\ D {\bf 63}, 063002 (2001),
  astro-ph/0005036.
  
  


  

  
      

\end{thebibliography}
\end{document}